\documentclass[aps,pra,twocolumn, superscriptaddress]{revtex4}
\usepackage{longtable}
\usepackage{graphicx}
\usepackage{dcolumn}
\usepackage{amsmath}
\usepackage{amssymb}
\usepackage{subfigure}
\usepackage{color,xcolor}
\usepackage{float}
\usepackage{epstopdf}

\def\beq{\begin{equation}}
\def\eeq{\end{equation}}

\begin{document}
\title{Semilocal and Hybrid Meta-Generalized Gradient Approximations Based on the Understanding of the Kinetic-Energy-Density Dependence}

\author{Jianwei Sun}
\affiliation{Department of Physics and Quantum Theory Group,
Tulane University, New Orleans, Louisiana 70118, USA}
\author{Robin Haunschild}
\affiliation{Department of Chemistry, Rice University, Houston, Texas 77005, USA} 
\author{Bing Xiao}
\affiliation{Department of Physics and Quantum Theory Group,
Tulane University, New Orleans, Louisiana 70118, USA} 
\author{Ireneusz W. Bulik}
\affiliation{Department of Chemistry, Rice University, Houston, Texas 77005, USA} 
\author{Gustavo E. Scuseria}
\affiliation{Department of Chemistry, Rice University, Houston, Texas 77005, USA}
\affiliation{Department of Physics and Astronomy, Rice University, Houston, Texas 77005, USA}
\affiliation{Chemistry Department, Faculty of Science, King Abdulaziz University, Jeddah 21589, Saudi Arabia}
\author{John P. Perdew}
\affiliation{Department of Physics and Quantum Theory Group, Tulane University, New Orleans, Louisiana 70118, USA}

\date{\today}

\begin{abstract}

We present a global hybrid meta-generalized gradient approximation (meta-GGA) with three empirical parameters, as well as its underlying semilocal meta-GGA and a meta-GGA with only one empirical parameter. All of them are based on the new meta-GGA resulting from the understanding of kinetic-energy-density dependence [J. Chem. Phys. {\bf 137}, 051101 (2012)]. The obtained functionals show robust performances on the considered molecular systems for the properties of heats of formation, barrier heights, and noncovalent interactions. The pair-wise additive dispersion corrections to the functionals are also presented.
\end{abstract}

\maketitle

\section{Introduction}
\label{sect:introduction}

The Kohn-Sham (KS) density functional theory (DFT) \cite{KS, Parr_Yang, PK} is one of the most widely used electronic structure theories for atoms, molecules and solids in various areas of physics, chemistry and molecular biology due to its computational efficiency and useful accuracy. It simplifies a many-electron wave-function problem to an auxiliary one-electron problem, with only its exchange-correlation part carrying the many-electron effects to be approximated in practice. Among numerous exchange-correlation approximations, the local spin density approximation (LSDA) \cite{KS, PW92, SPS_PRB_2010}, the standard Perdew-Burke-Ernzerhof (PBE) generalized gradient approximation (GGA) \cite{PBE}, and the Becke-3-Lee-Yang-Parr (B3LYP)~\cite{Becke_JCP_1993,B88,LYP,B3} hybrid GGA dominate the user market of DFT~\cite{Burke_JCP_2012}. The former two are efficient semilocal functionals widely used for extended systems, while the latter is a computationally more expensive nonlocal functional that hybridizes a GGA with the exact exchange energy and is popular for finite systems. At the semilocal level, however, meta-GGA (MGGA) is the highest rung of the so-called Jacob's ladder in DFT \cite{Jacob_Ladder} and potentially the most accurate one \cite{SMRKP}, which can also serve as a better base for hybridizing with the exact exchange energy.

Semilocal approximations (e.g., Refs.~\cite{PW92, SPS_PRB_2010, PBE, PBEsol, TPSS, PRCCS}) of the form
\beq
E_{\rm xc}^{\rm sl}[n_{\uparrow}, n_{\downarrow}]=\int d^3r n \epsilon_{\rm xc}^{\rm sl}(n_{\uparrow},n_{\downarrow}, \nabla n_{\uparrow},\nabla n_{\downarrow},\tau_{\uparrow},\tau_{\downarrow})
\label{semi_local}
\eeq
require only a single integral over real space and so are practical even for large molecules or unit cells. In Eq. \eqref{semi_local}, $n_{\uparrow}$ and $n_{\downarrow}$ are the electron densities of spin $\uparrow$ and $\downarrow$, respectively, $\nabla n_{\uparrow,\downarrow}$ the local gradients of the spin densities, $\tau_{\uparrow,\downarrow}=\sum_k \left |\nabla \psi_{k{\uparrow,\downarrow}} \right |^2/2$ the kinetic energy densities of the occupied KS orbitals $\psi_{k\sigma}$ of spin $\sigma$ ($\sigma = \uparrow,\downarrow$), and $ \epsilon_{\rm xc}^{\rm sl}$ the approximate exchange-correlation energy per electron. All equations are in atomic units. 
In addition to $n_{\uparrow}$ and $n_{\downarrow}$, the only ingredients in LSDA, GGAs also use the density gradients; MGGAs additionally include the kinetic energy densities $\tau_{\sigma}$. With the encoded information of shell structures from the kinetic energy densities, MGGAs can distinguish different orbital-overlap regions \cite{SXR_JCP_2012}, and thus deliver simultaneous accurate ground-state properties for molecules, surfaces, and solids -- which couldn't be obtained with a GGA or LSDA \cite{SMRKP}. Computationally, MGGAs are not much more expensive than LSDA or GGA~\cite{SSTP_JCP_2003,FP_JCP_2006}. In computations for molecules containing transition-metal atoms, the Tao-Perdew-Staroverov-Scuseria (TPSS) \cite{TPSS} MGGA is only $30\%$ slower \cite{FP_JCP_2006} than PBE.

To understand the performance of MGGAs, a recent study~\cite{SXR_JCP_2012} investigated the effect of $\tau$-dependence on MGGAs through the inhomogeneity parameter $\alpha=(\tau-\tau^W)/\tau^{\rm unif}$. Here, $\tau=\sum_\sigma \tau_\sigma$, $\tau^W=\frac{1}{8}|\nabla n|^2/n$ is the  von Weizs$\ddot{{\rm a}}$cker kinetic energy density, and $\tau^{\rm unif}=\frac{3}{10}(3 \pi^2)^{2/3}n^{5/3}$ is the kinetic energy density of the uniform electron gas (UEG). 
$\tau^W$ is a lower bound on $\tau$ with $\tau^W=\tau$ only for a single-orbital region~\cite{KPB_IJQC_1999}. Therefore, $\alpha \geqslant 0$ measures locally how far $n({\bf r})$ deviates from being of single-orbital character on the scale of the UEG, and thus characterizes the extent of orbital overlap. Besides the ability of $\alpha$ to distinguish different orbital-overlap regions, Ref.~\onlinecite{SXR_JCP_2012} further showed that the effect of the $\alpha$-dependence on MGGAs is qualitatively equivalent to that of the dependence on the reduced density gradient $s=|\nabla n|/[2(3 \pi ^2)^{1/3} n ^{4/3}]$, another dimensionless inhomogeneity parameter that measures how fast and how much the density varies on the scale of the local Fermi wavelength $2\pi/k_F$ with $k_F=(3\pi^2n)^{\rm 1/3}$. The $s$-dependence is well understood at the GGA level~\cite{PBE, PBEsol}, which MGGAs inherit~\cite{TPSS, PRCCS, SXR_JCP_2012}.

As a result of the study on the $\tau$-dependence~\cite{SXR_JCP_2012}, a simple exchange functional, where the $\alpha$-dependence is disentangled from the $s$-dependence, was constructed as an interpolation between the single-orbital regime ($\alpha=0$) and the slowly-varying density regime ($\alpha \approx 1$)~\cite{SXR_JCP_2012}, which we will discuss more in Section~\ref{sect:construction}. In analogy to the monotonically increasing $s-$dependence that GGAs are often designed to have to favor more inhomogeneous electron densities, the simple exchange functional has a monotonically decreasing $\alpha-$dependence to favor more the single-orbital regions. When combined with a variant of the PBE correlation ~\cite{PRCCS, SXR_JCP_2012}, the resulting MGGAs (denoted as MGGA$_-$MS with MS standing for "made simple") \cite{SXR_JCP_2012} performs equally well for atoms, molecules, surfaces, and solids, with an overall performance that is comparable to the sophisticated revised TPSS (revTPSS) MGGA \cite{PRCCS}. Moreover, MGGA\_MS yields excellent binding energies for the W6 water clusters~\cite{SXR_JCP_2012}, even better than those from the highly parametrized M06-L MGGA \cite{ZT_JCP_2006} whose training sets include noncovalent interactions (hydrogen bonds and van der Waals interactions). Further tests on general main group thermochemistry, kinetic, and noncovalent interactions~\cite{HSXRCTP_JCTC_2012} showed that MGGA\_MS, though not as good as M06-L, systematically improves the descriptions for the noncovalent interactions over conventional semilocal functionals, e.g., PBE, TPSS, and revTPSS.

By including training sets of noncovalent interactions, the M06-L MGGA was trained to capture medium-range exchange and correlation energies that dominate equilibrium structures of noncovalent complexes \cite{ZT_JCP_2006}. For example, M06-L delivers good binding energies for the S22 set that is dominated by noncovalent interactions and will be discussed more in Sec.~\ref{sect:results}, good structural and energetical properties of layered materials bound by van der Waals interactions (e.g., graphite, hexagonal boron nitride, and molybdenum disulfite~\cite{MFH_JPCL_2010}), and good adsorption energies of aromatic molecules adsorbed on coinic metal surfaces~\cite{FMH_JCP_2011}. 
M06-L represents one of the developments of MGGAs that are fitted to a variety of training sets with numerous parameters to try to improve different properties together~\cite{ZT_JCP_2006, PT_JPCL_2012, VSXC}. However, it should be stressed that too many fitting parameters could cause problems, e.g., convergence of calculations and the related smoothness of binding curves~\cite{GG11}, and the prediction for systems far from training sets might not be realistic~\cite{SRHMB_PRL_2012}.   

Semilocal (sl) approximations can be reasonably accurate for the near-equilibrium and compressed ground-state properties of ‘‘ordinary’’ matter, where neither strong correlation nor long-range van der Waals interaction is important. However, stretched bonds that arise in transition states of chemical reactions or in the dissociation limits of some radical or heteronuclear molecules, etc., require full nonlocality~\cite{PSTS_PRA_2008}. The global hybrid functionals discussed later can provide such needed nonlocality. Long-range van der Waals interactions, originated from spontaneous correlations between two distant electron densities, also require full nonlocality~\cite{DRSLL_PRL_2004}, which is different from the previous one~\cite{RPC_JCTC_2010}. To obtain long-range van der Waals interactions, there are various DFT-based dispersion techniques, including the post-DFT empirical pairwise potential corrections~\cite{WY_JCP_2002,Becke_JCP_2005,BJ_JCP_2005,JB_JCP_2005,BJ_JCP_2005_2,Grimme_JCC_2006,GAEK_JCP_2010,TS_PRL_2009} and the van der Waals density functionals (vdW-DFs)~\cite{DRSLL_PRL_2004, VV_PRL_2009, KBM_JPCM_2010}, which are based on the pairwise additivity and problematic for metallic systems~\cite{DWR_PRL_2006, RPTCP_PRL}. 
See Ref.~\onlinecite{KM_JCP_2012} for an overview of the progess on DFT-based dispersion techniques. Conventional semilocal functionals, e.g., the widely used PBE GGA, even lack the ability to describe noncovalent interactions near equilibrium, while the M06-L MGGA and MGGA\_MS as well as its variant proposed here improve over PBE. 

The global hybrid (gh) idea, due to Becke~\cite{Becke_JCP_1993, Becke_JCP_1993_2}, introduces some full nonlocality into the calculation but only at the level of $E_x^{\rm exact}$, which can be evaluated semianalytically from the Kohn-Sham orbitals in some computer codes. In its simplest (one parameter) version~\cite{PBE0}:
\beq
E_{\rm xc}^{\rm gh}=aE_x^{\rm exact}+(1-a)E_x^{\rm sl}+E_c^{\rm sl}
\label{global_hybrid}
\eeq
where the exact-exchange mixing parameter $a$ takes an empirical value in the range $0 \le a \le 1$. A rough but convincing argument \cite{CPR_JCTC_2010} for why Eq. (\ref{global_hybrid}) works is: semilocal exchange-correlation typically overestimates atomization energies and underestimates energy
barriers of chemical reactions, while exact exchange without correlation makes errors of the opposite sign, so that mixing of the two will yield better atomization energies and barriers than either alone. Therefore, values of $a$ vary with the choice of semilocal functionals when fitting Eq. (\ref{global_hybrid}) to formation enthalpies. The more a semilocal functional overbinds molecules, the larger the value of $a$ is. Csonka {\it et al.}~\cite{CPR_JCTC_2010} found $a$ to be 0.60 for the PBEsol GGA, 0.32 for the PBE GGA, and 0.1 for the TPSS and revTPSS MGGAs. Improving the underlying semilocal functional reduces the value of $a$ needed to fit the atomization energies or enthalpies of formation, which however can worsen the barrier heights. Some global hybrids~\cite{Becke_JCP_1997, ZT_ACR_2008} thus employ highly fitted semilocal parts intended not to be accurate by themselves but to work well with a fraction of exact exchange.

The popular B3LYP~\cite{Becke_JCP_1993, B88,LYP,B3} hybrid GGA has a somewhat more complicated form and reads:
\begin{eqnarray}
E_{\rm xc}^{\rm B3LYP}&=&aE_x^{\rm exact}+(1-a)E_x^{\rm LSDA}+b\Delta E_x^{\rm B88} + \nonumber \\ && +(1-c)E_c^{\rm VWN3}+cE_c^{\rm LYP}
\label{B3LYP}
\end{eqnarray}

where $E_c^{\rm VWN3}$~\cite{VWN_CJP_1980} is the LSDA correlation energy in the random phase approximation which does not yield the correct uniform electron gas limit. In the equation, a=0.2, b=0.72, and c=0.81. If we use the similar idea as in Eq.~\eqref{global_hybrid} to define the underlying semilocal functional for B3LYP by replacing $E_x^{\rm exact}$ with $E_x^{\rm B88}$, we end up with a GGA, denoted as BLYP$^*$, $E_{\rm xc}^{\rm BLYP^*}=E_x^{\rm LSDA}+(a+b)\Delta E_x^{\rm B88}+(1-c)E_c^{\rm VWN3}+cE_c^{\rm LYP}$. BLYP$^*$ differs from BLYP in reducing the gradient correction of the B88 exchange to (a+b)=92\% and mixing the VWN3 and LYP correlations. This difference confirms the rough argument of the previous paragraph. Typically, BLYP underbinds molecules, and inclusion of exact exchange worsens the underbinding. The reduction of the gradient correction of the B88 exchange and the mixture of the VWN3 and LYP correlations then turn BLYP$^*$ into an overbinding functional suitable for hybridization with the exact exchange. The idea of tuning the gradient correction of B3LYP can be transfered to the construction of our hybrid MGGA, where both the $s-$dependence and the $\alpha-$dependence are tuned.  


In the next section, we will present the computational details, followed by the constructions of semilocal and hybrid MGGAs in Section \ref{sect:construction}. Results and discussions will be given in Section \ref{sect:results}. And we give the conclusions in Section \ref{sect:conclusions}.

\section{Computational details}
\label{sect:details}

The functionals described here were implemented in the
development version of the \mbox{\sc Gaussian} electronic structure program \cite{GDV-G1}.
All calculations employ the fully uncontracted 6-311++G($3df,3pd$) basis set
\cite{6311ppG3df3pd-1, 6311ppG3df3pd-2} to obtain benchmark quality results.
We used the {\tt UltraFine} grid with 99 radial shells and 590 angular points for numerical integration of the DFT XC potential. 
The \mbox{\sc Gaussian} program was used for all calculations presented in this study.

Different training sets were used to adjust empirical parameters.
For the AE6 and BH6 tests sets \cite{AE6BH6} we employed reference data from highly accurate
quantum-chemical calculations \cite{AE6BH6Ref}.
Experimental reference values were used for G2/97 (148 molecules) \cite{G2-1, G2-2} and
BH42/03 (21 forward and reverse hydrogen transfer barrier heights) \cite{BH42}.
The fitted parameters would not change much if the highly accurate quantum-chemical reference data from Ref. \onlinecite{G2Ref}
would have been used.

The superset of G2/97 and G3-3 (75 molecules) \cite{G3-3} (a total of 223 molecules comprising the G3/99 test set)
\nocite{TPSSh} \cite{fn1}, the S22 test set for weak interactions \cite{S22ref}, as well as the barrier height test sets HTBH38/04
(19 forward and reverse hydrogen transfer barrier heights) \cite{BH42} and NHTBH38/04
(19 forward and reverse non-hydrogen transfer barrier heights) \cite{NHTBH38} are employed
for assessment of the fitted functionals.
The employed geometries and reference values are available from the Supporting Information of Refs. \onlinecite{Lh-PBE} and
\onlinecite{S22ref}.
The counterpoise correction \cite{CP} was used to reduce the basis set superposition errors for calculations of weak interactions.
A vibrational scale factor of 0.9854 was used for calculations of heats of formation on the B3LYP/6-31G($2df,p$) level of theory
as recommended in Ref. \onlinecite{G3geom}.

The performance of our fitted functionals will be compared to related and popular density functionals:
PBE \cite{PBE}, TPSS \cite{TPSS}, revTPSS \cite{PRCCS}, M06-L \cite{ZT_JCP_2006}, MGGA\_MS0 \cite{SXR_JCP_2012},
PBEh \cite{PBE0}, TPSSh \cite{TPSSh}, revTPSSh \cite{PRCCS}, B3LYP \cite{B88,LYP,B3}, BLYP \cite{B88,LYP}, and M06 \cite{M06}.
Calculations of open-shell species were carried out via spin-unrestricted formalisms.
Errors are reported as $calculated - reference$.

\section{Construction of semilocal and hybrid meta-GGAs}
\label{sect:construction}

The semilocal exchange energy of a spin-unpolarized density can be written as:
\beq
E_x^{\rm sl}[n]=\int d^3 r n \epsilon_x^{\rm unif}(n)F_x(p, \alpha).
\label{eq:Ex_sl}
\eeq
Here, $\epsilon_x^{\rm unif}(n)=-\frac{3}{4\pi}(\frac{9\pi}{4})^{1/3} / r_s$ is the exchange energy per particle of a UEG with $r_s=(4 \pi n /3)^{-1/3}$, $p=s^2$, and $F_x$ the enhancement factor with $F_x=1$ for LSDA. The expression for the spin-polarized case follows from the exact spin scaling~ \cite{PKZB}. In Ref.~\onlinecite{SXR_JCP_2012}, a simple exchange enhancement factor that disentangles the $\alpha$- and $p$-dependence was proposed,
\beq
F_x^{\rm int}(p, \alpha)=F_x^{\rm 1}(p)+f(\alpha)[F_x^{\rm 0}(p)-F_x^{\rm 1}(p)],
\label{eq:Fx_MGGA}
\eeq
where $F_x^{\rm 1}(p)=F_x^{\rm int}(p, \alpha=1)=1+\kappa-\kappa/(1+\mu^{\rm GE}p/\kappa)$ and $F_x^{\rm 0}(p)=F_x^{\rm int}(p, \alpha=0)=1+\kappa-\kappa/(1+(\mu^{\rm GE}p+c)/\kappa)$. $F_x^{\rm int}(p, \alpha)$ interpolates between $F_x^{\rm 0}(p)$ and $F_x^{\rm 1}(p)$ through $f(\alpha)$. Here, in order to tune the $\alpha$ dependence, we add a parameter $b$ in the denominator of the original $f(\alpha)$, which reads now $f(\alpha)=(1-\alpha^2)^3/(1+\alpha^3+b \alpha^6)$.  For a slowly varying density where $\alpha \approx 1$, $f(\alpha)$ is controlled by its numerator and we recover the second order gradient expansion with the first principle coefficient $\mu^{\rm GE}=10/81$~\cite{AK_PRB_1985} as in Ref.~\onlinecite{SXR_JCP_2012}. The parameter $c$ is determined for each $\kappa$ to reproduce the exchange energy of the hydrogen atom, where $\alpha=0$. Since $b\alpha^6$ is negligible for $0 < \alpha < 1$ and the numerator of $f(\alpha)$ modulates the behavior of $F_x^{\rm int}(p, \alpha)$ for $\alpha \approx 1$, $b$ mainly controls the behavior for $\alpha>1$. On the other hand, $\kappa$ mainly controls the behavior of $F_x^{\rm int}(p, \alpha)$ for large $s$. In the original form~\cite{SXR_JCP_2012}, $b=1$, and $\kappa=0.29$ was determined by the Hartree-Fock exchange energy of the 12-noninteracting-electron hydrogenic anion with nuclear charge Z=1~\cite{SSPTD_PRA_2004}. Fig. 2 of Ref.~\onlinecite{SXR_JCP_2012} showed that the shell regions of the hydrogenic anion are associated with $\alpha<1$ and the intershell regions with $\alpha>1$, which thus can provide information to guide the exchange functional from $\alpha \approx 1$ to $\alpha \to \infty$. Here, we release the constraint of the hydrogenic anion and rely on some molecular training sets to provide information to determine both $b$ and $\kappa$ by fitting them to heats of formation and barrier heights. For the correlation part, we use the variant of the PBE correlation \cite{PRCCS, SXR_JCP_2012}.


When we plug the above exchange and correlation parts into Eq. (\ref{global_hybrid}), we obtain a global hybrid MGGA with three parameters, $\kappa$, $b$, and $a$, which control the dependence of the global hybrid on the density gradient, the kinetic energy density, and the exact exchange, respectively.
In the next section, we describe the parametrization procedure and the performance of our best functionals.

\section{Results and Discussions}

\label{sect:results}

\begin{figure}
\includegraphics[width=\linewidth]{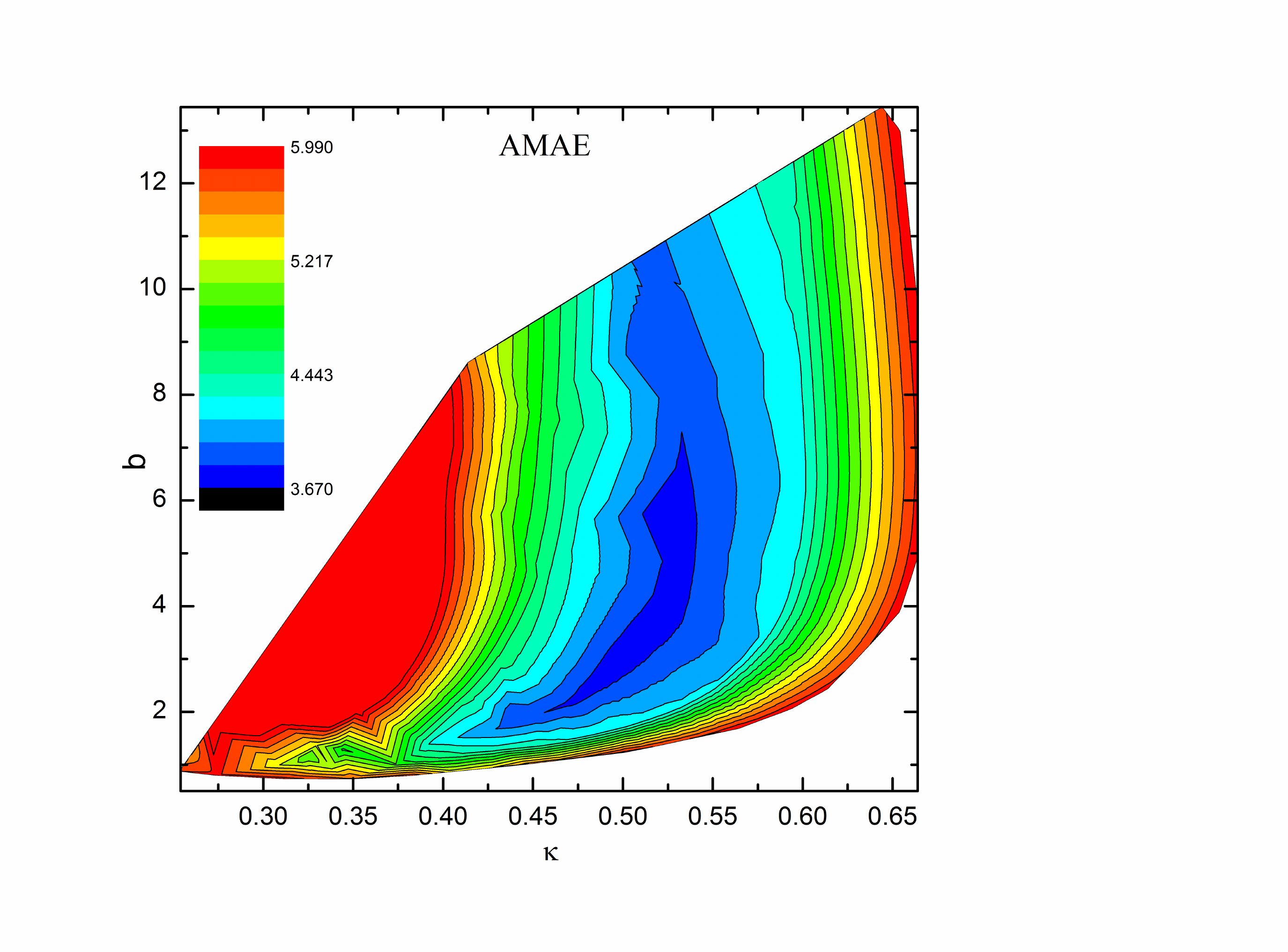}
\caption{The AMAE surface for the semilocal MGGA\_MS$i$ variants with errors in kcal/mol shows a broad range of rather small errors}
\label{figure:AMAE}
\end{figure}

To get a broad overview about the errors which the semilocal functional class MGGA\_MS makes using different values of
$\kappa$ and $b$, we define the average mean absolute error (AMAE) using the mean absolute error (MAE) values from the AE6 and BH6
test sets:
\begin{eqnarray}
\label{AMAE}
{\rm AMAE} & = & 0.5 \left[ {\rm MAE(AE6)} + {\rm MAE(BH6)}\right] ,
\end{eqnarray}
and present the plotted surface of AMAE values in figure \ref{figure:AMAE}.
The figure shows that there are two useful possibilities of fitting a semilocal functional of our
MGGA\_MS form: First, we can design a one-parameter functional by setting the parameter $b$ to one
and fitting $\kappa$ to minimize a chosen error measure. Second, both parameters can be fitted independently.
The second choice will yield a lower AMAE value.
For the independent fit, we also include the exact exchange admixture parameter $a$ into the fitting procedure.
The corresponding semilocal functional will be obtained by setting $a$ to zero while keeping the other
global hybrid parameters fixed. 

Our chosen error measure contains the mean absolute errors (MAE's) of the
heats of formation of the G2/97 data set and barrier heights of the BH42/03 test sets:
\begin{eqnarray}
SW = 0.5 \left[ \text{MAE(G2/97) + MAE(BH42/03)} \right] .
\end{eqnarray}
The resulting parameter sets, ME, and MAE values are shown in table \ref{table:Parameter_G2BH42}
and compared to other popular semilocal and hybrid functionals.
MGGA\_MS0 is the original MGGA\_MS functional from Ref. \onlinecite{SXR_JCP_2012}
with no parameter fitting to training sets while MGGA\_MS1 is our semilocal one-parameter functional.
MGGA\_MS2h is the new hybrid functional and MGGA\_MS2 its semilocal complement.
Our hybrid MGGA\_MS2h has a rather low exact exchange admixture of 9\% as expected when a good semilocal
functional is hybridized~\cite{CPR_JCTC_2010}.
Comparing the parameter set of MGGA\_MS2 with the AMAE surface in figure \ref{figure:AMAE} shows that its parameter set is quite
close to the global minimum of the AMAE, although it has not been refitted after exact exchange admixture was reduced
from 9\% to 0\%.
The functional BLYP* is not commonly used, but it is the underlying semilocal functional of B3LYP. 
It is included in our tables for comparison purposes only. As shown in tables \ref{table:Parameter_G2BH42} and \ref{table:G3BH76}, BLYP* turns the underestimations of BLYP for atomization energies into overestimations.
From all semilocal functionals in table \ref{table:Parameter_G2BH42}, MGGA\_MS1 performs best for G2/97
while M06-L performs best for BH42/03.
MGGA\_MS0, MGGA\_MS1, and MGGA\_MS2 functionals are the next best performers for the BH42/03 test set,
which behave similarly despite very different parameter sets.
For the hybrids, B3LYP performs best for G2/97 and M06 for BH42/03.
Our global hybrid MGGA\_MS2h performs similarly to B3LYP and PBEh for the BH42/03 test set and similar to PBEh for the
G2/97 test set.
Note 
that PBE, TPSS, revTPSS, and MGGA\_MS0 are nonempirical
functionals, while the number of empirical parameters -- as indicated in the parentheses --
of MGGA\_MS1 (1), MGGA\_MS2 (2), TPSSh (1), revTPSSh (1),
and MGGA\_MS2h (3) are an order of magnitude smaller than those of M06-L and M06.

MGGA\_MS1 performs better than MGGA\_MS2 and MGGA\_MS0 for all molecular test sets discussed so far.
This and figure \ref{figure:AMAE} indicate that one can construct another MGGA\_MS functional which performs even better
than MGGA\_MS1. However, figure \ref{figure:AMAE} suggests that the improvement over MGGA\_MS1 will not be significant.
We found that $\kappa = 0.514$, $c = 0.14352$, and $b = 2.0$ is close to optimal for the error measure $SW$, but the minimum in
parameter space is rather shallow and broad.
With MAE values for G2/97 of 4.5 kcal/mol and for BH42/03 of 5.6 kcal/mol, this variant would improve over MGGA\_MS1
only slightly for G2/97 and not at all for BH42/03 on the expense of an additional empirical parameter.
Thus, we drop this variant from now on.

%
\begin{center}
\noindent
\begin{table*}[htb]
\caption{Selected parameters, mean errors (ME), and mean absolute errors (MAE) of semilocal and hybrid functionals for the G2/97 and BH42/03 test sets. The ME and MAE values are in kcal/mol.}
\centering
\vspace{2mm}
\begin{tabular}{l c c c c c c c c c c}
\hline\hline
            &  $a$  & $\kappa$ &   $c$   &  $b$  & &  \multicolumn{2}{c}{G2/97} & & \multicolumn{2}{c}{BH42/03} \\
\cline{7-8} \cline{10-11}
            &       &          &         &       & & ME & MAE                   & & ME & MAE           \\
\hline
PBE         & 0.00  & 0.804       &   --    & --    & & -16.1 & 16.9               & & -9.7 & 9.7   \\
BLYP        & 0.00  & --       &   --    & --    & & -0.6  &  7.4               & & -8.0 & 8.0 \\
BLYP*       & 0.00  & --       &   --    & --    & & -8.1  &  9.0               & & -8.4 & 8.4   \\
TPSS        & 0.00  & 0.804    &   --    & --    & & -5.7  &  6.4               & & -8.4 & 8.4   \\
revTPSS     & 0.00  & 0.804    &   --    & --    & & -4.7  &  5.7               & & -7.7 & 7.7   \\
M06-L       & 0.00  & --       &   --    & --    & & -3.0  &  5.1               & & -4.4 & 4.4   \\
MGGA\_MS0   & 0.00  & 0.29     & 0.28771 & 1.0   & & -0.6  &  6.8               & & -5.9 & 6.0   \\
MGGA\_MS1   & 0.00  & 0.404    & 0.18150 & 1.0   & &  2.3  &  4.9               & & -5.5 & 5.6   \\
MGGA\_MS2   & 0.00  & 0.504    & 0.14601 & 4.0   & & -0.8  &  5.1               & & -5.8 & 5.8   \\
\hline
PBEh        & 0.25  & 0.804       &   --    & --    & & -2.6  & 5.0                & & -4.7 &  4.7  \\
TPSSh       & 0.10  & 0.804    &   --    & --    & & -1.9  & 4.4                & & -6.6 &  6.6  \\
revTPSSh    & 0.10  & 0.804    &   --    & --    & & -1.2  & 4.5                & & -6.0 &  6.0  \\
B3LYP       & 0.20  & --       &   --    & --    & &  0.9  & 3.1                & & -4.7 &  4.7  \\
M06         & 0.27  & --       &   --    & --    & & -3.4  & 4.1                & & -2.3 &  2.3  \\
MGGA\_MS2h  & 0.09  & 0.504    & 0.14601 & 4.0   & &  2.7  & 4.9                & & -4.5 &  4.6  \\
\hline\hline
\end{tabular}
\label{table:Parameter_G2BH42}
\end{table*}
\end{center}

\begin{center}
\noindent
\begin{table*}[htb]
\caption{Mean errors (ME), and mean absolute errors (MAE) of semilocal and hybrid functionals for the G3-3, G3/99, HTBH38, and NHTBH38 test sets in kcal/mol}
\begin{tabular}{l ccccccccccc}
\hline\hline
            & \multicolumn{2}{c}{G3-3} & & \multicolumn{2}{c}{G3/99} & & \multicolumn{2}{c}{HTBH38/04} & & \multicolumn{2}{c}{NHTBH38/04}\\
\cline{2-3} \cline{5-6} \cline{8-9} \cline{11-12}
            &    ME      &    MAE      & &    ME       &    MAE      & &   ME        &   MAE        & &    ME        &  MAE       \\
\hline
PBE         &    -32.6   &    32.6     & &    -21.6    &   22.2      & &   -9.7      &  9.7         & & -8.5         & 8.6        \\
BLYP        &     12.5   &    14.1     & &      3.8    &    9.7      & &   -7.9      &  7.9         & & -8.7         & 8.8        \\
BLYP*       &     -5.7   &     8.6     & &     -7.3    &    8.8      & &   -8.4      &  8.4         & & -8.7         & 8.8        \\
TPSS        &     -6.4   &     6.5     & &     -6.0    &    6.5      & &   -8.2      &  8.2         & & -9.0         & 9.1        \\
revTPSS     &     -4.2   &     4.7     & &     -4.5    &    5.4      & &   -7.4      &  7.4         & & -9.3         & 9.3        \\
M06-L       &     -4.2   &     7.4     & &     -3.4    &    5.8      & &   -4.5      &  4.5         & & -3.2         & 3.7        \\
MGGA\_MS0   &     -5.9   &    11.8     & &     -2.4    &    8.5      & &   -5.6      &  5.7         & & -6.4         & 6.9        \\
MGGA\_MS1   &      3.2   &     5.2     & &      2.6    &    5.0      & &   -5.4      &  5.5         & & -6.3         & 6.7        \\
MGGA\_MS2   &     -4.1   &     7.6     & &     -1.9    &    6.0      & &   -5.7      &  5.7         & & -7.0         & 7.3        \\
\hline
PBEh        &     -9.6   &   10.4      & &     -4.9    &    6.8      & &   -4.7      &  4.7         & & -3.2         & 3.7        \\
TPSSh       &     -1.0   &    3.5      & &     -1.6    &    4.1      & &   -6.4      &  6.4         & & -6.9         & 7.0        \\
revTPSSh    &      0.6   &    3.6      & &     -0.6    &    4.2      & &   -5.8      &  5.8         & & -7.1         & 7.2        \\
B3LYP       &      7.9   &    8.2      & &      3.3    &    4.8      & &   -4.5      &  4.6         & & -4.6         & 4.7        \\
M06         &     -4.4   &    5.4      & &     -3.7    &    4.5      & &   -2.3      &  2.3         & & -1.7         & 2.2        \\
MGGA\_MS2h  &      1.5   &    5.1      & &      2.3    &    4.9      & &   -4.4      &  4.4         & & -5.3         & 5.7        \\
\hline\hline
\end{tabular}
\label{table:G3BH76}
\end{table*}
\end{center}

\begin{figure}
\subfigure[]{\label{figure:Fx_alpha}\includegraphics[width=\linewidth]{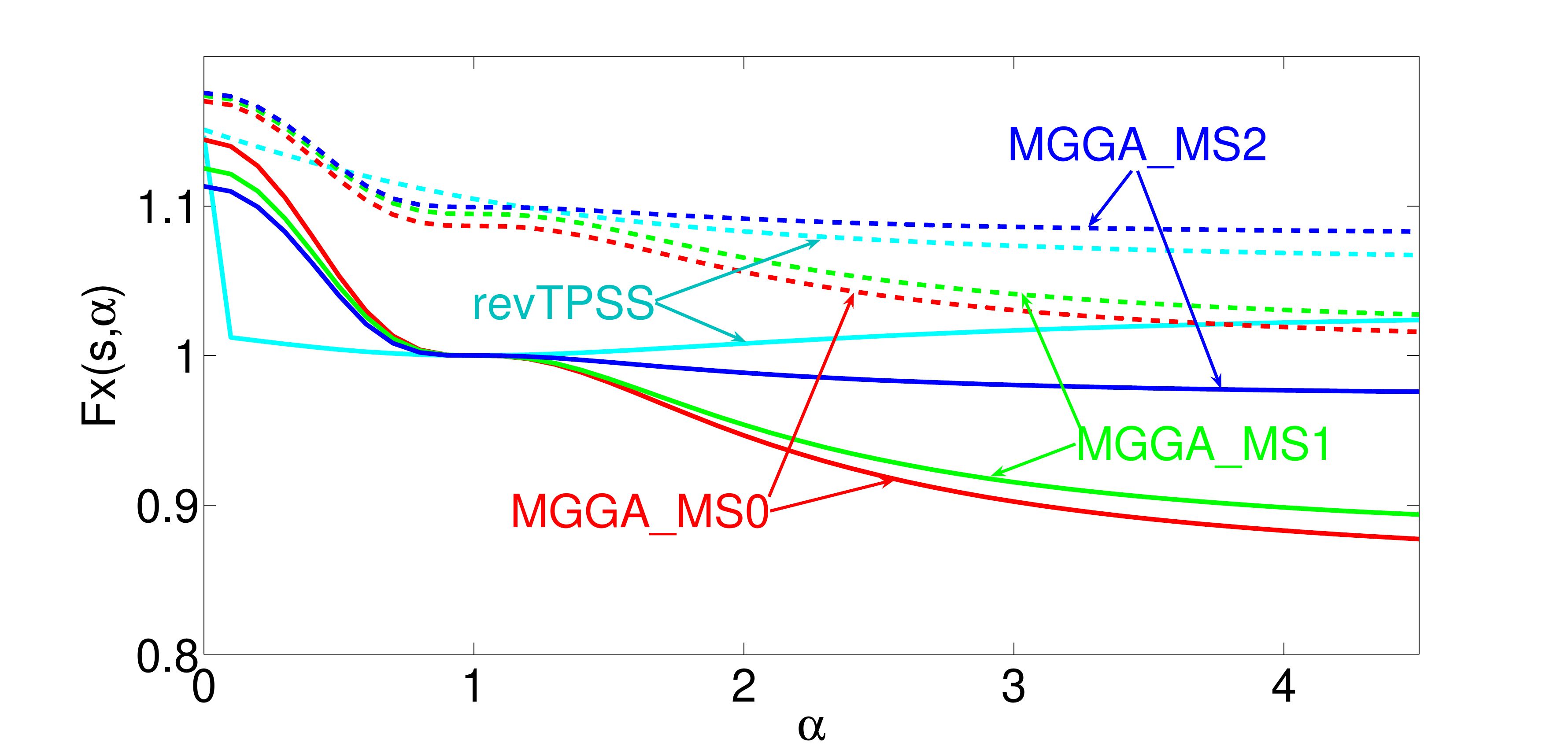}}
\\ 
\subfigure[]{\label{figure:Fx_s}\includegraphics[width=\linewidth]{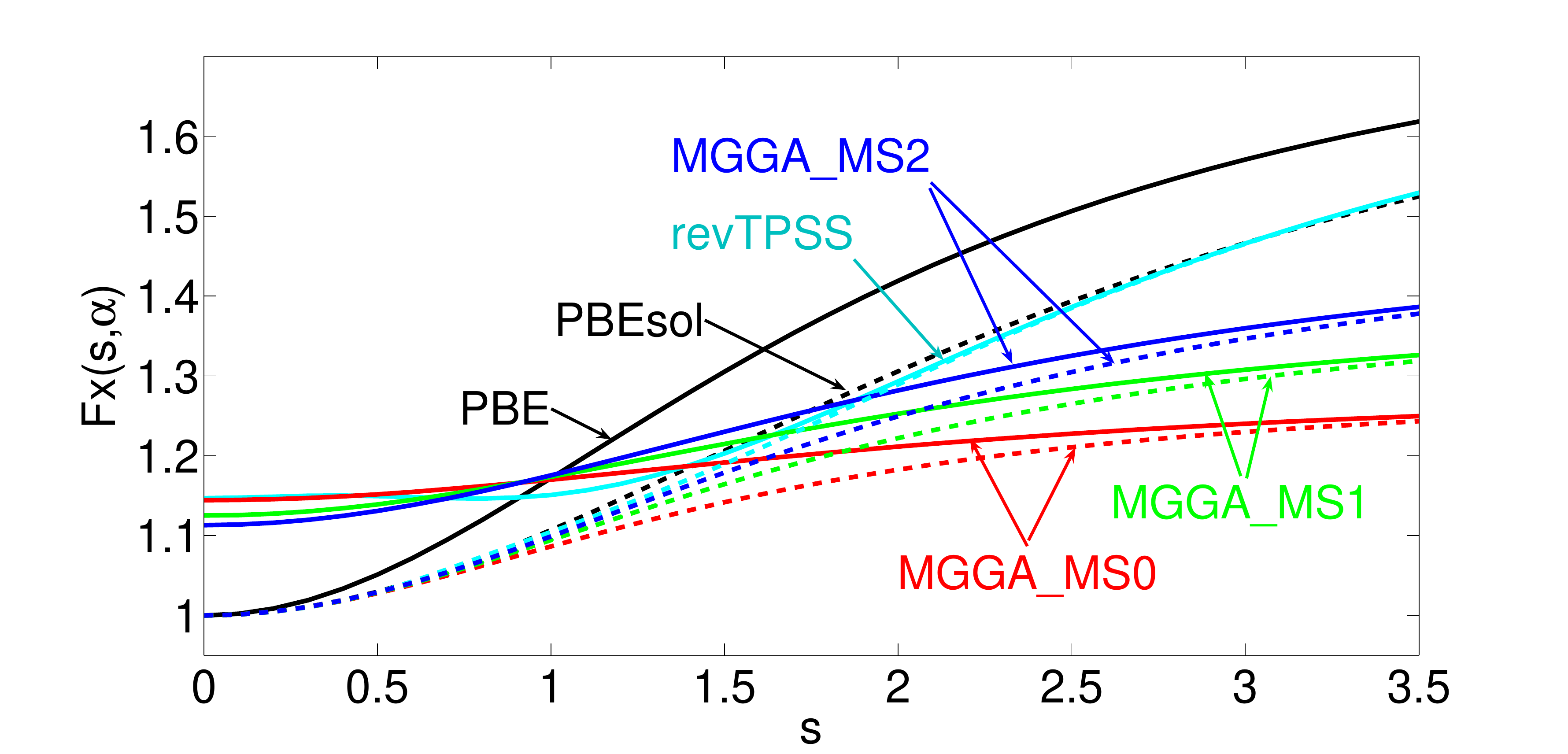}}
\caption{(a) Exchange enhancement factors vs. $\alpha$ for different $s$. The solid curves are for $s=0.001$ and dashed for $s=1.001$.  (b) Exchange enhancement factors vs. $s$ for different $\alpha$. In addition to the PBE and PBEsol curves, the solid curves are for $\alpha=0$ and dashed for $\alpha=1$.}
\label{figure:Fx_alpha_s}
\end{figure}

To test the deviation from the exact exchange energy of the 12-noninteracting-electron hydrogenic anion, we calculate the exchange energies for the new parameter sets of (b, $\kappa$), which are -1.8562 Ha for (b=1, $\kappa=0.404$) of MGGA\_MS1 and -1.8558 Ha for (b=4, $\kappa=0.504$) of MGGA\_MS2, respectively.
In comparison with the exact one, -1.8596 Ha, the deviations are small, which indicates that the information of the hydrogenic anion is preserved by Eq. \ref{eq:Fx_MGGA} and complemented by that from training sets. Figure \ref{figure:Fx_alpha_s} shows the exchange enhancement factors for different $\alpha$ and $s$ values of our new functionals and compares them to the revTPSS functional. Figure \ref{figure:Fx_alpha} shows revTPSS has the order of limits anomaly at small $s$ and small $\alpha$ ~\cite{TPSS, PRCCS}, which can be eliminated by regularizing revTPSS at these regions~\cite{regTPSS}. The family of MGGA\_MS doesn't have such a problem by construction. The enhancement factors of different MGGAs considered here are on top of each other around $\alpha \approx 1$ at the curves of $s=0.001$, due to the constraint of the second order gradient expansion. Figure \ref{figure:Fx_s} shows that each MGGA is insensitive to $\alpha$ for large $s$. For the $\alpha=0$ curves, which is modulated by the constraint of the exchange energy of the hydrogen atom, $F_x$ gets smaller for small $s$ regions when a larger $\kappa$ is used. The behavior of these MGGAs at large s is determined by the value of $\kappa$. The larger $\kappa$, the larger enhancement factor at large $s$. Due to the qualitative equivalence between the $s-$ and $\alpha-$dependence~\cite{SXR_JCP_2012}, we observe for different MGGAs that the faster the increase of F$_x$ with $s$, the slower the decrease of F$_x$ with $\alpha$.

It is not too surprising that our functionals fitted to G2/97 and BH42/03 are among the best performers for these test sets.
Table \ref{table:G3BH76} shows the performance for the heats of formation of the test sets G3-3 and G3/99 as well as
the barrier heights of the data sets HTBH38/04 and NHTBH38/04.
The Minnesota functionals perform best for the barrier heights,
which is not too surprising as both barrier height test sets are contained in the training set of M06 and M06-L.
The next best semilocal functional for both barrier height test sets is MGGA\_MS1.
Among the global hybrids, PBEh and B3LYP perform better than MGGA\_MS2h for the NHTBH38/04 test set
but not for the HTBH38/04 test set which was part of the training set.
TPSSh performs best for G3-3 and G3/99 within our choice of global hybrids.
Interestingly, all global hybrids in table \ref{table:G3BH76} (except PBEh) perform very similar on average for the G3/99 test set.
Among the semilocal functionals revTPSS performs best for G3-3 and MGGA\_MS1 for G3/99.

Aside from heats of formation and barrier heights, we would like to test our functionals for a property which was
not part of the training set. We choose the weak interactions of the S22 test set.
Table \ref{table:S22} presents the ME and MAE values for the S22 test set and its subsets WI8 (including 8 dispersion-bound complexes), MI7 (including 7 mixed complexes), and HB7 (including 7 hydrogen-bonded complexes).
Before discussing the results, we would like to test the convergence of MGGAs with respect to the integration grid for dispersion-bound complexes. Johnson {\it et al}~\cite{JBSD_JCP_2009} found that very large integration grids (much finer than the {\tt UltraFine}) are required to remove oscillations in potential energy surfaces (PES) for dispersion-bound complexes for most of the MGGAs they studied. For the chosen simple dispersion-bound NeAr, we confirmed that the M06L MGGA, which was in Johnson's test list, needs a very large grid to yield a smooth PES. However, our MGGAs converge much faster for the total energies than M06L does, and yield converged smooth curves with the {\tt UltraFine} grid.
The Minnesota functionals, which have been fitted for weak interactions, perform best for all these data sets of weak interactions.
Among semilocal functionals, MGGA\_MS0 is the second best performer for the S22 set and its subsets. MGGA\_MS2 on average is comparable to MGGA\_MS0 for the WI8 and MI7 subsets, but worse for the HB7 subset, while MGGA\_MS1 loses the accuracy for the weak interactions compared to the other two in the MGGA\_MS family. As expected for WI8 and MI7, where dispersion interactions are important, PBE, TPSS, and revTPSS yield large MAEs, which are worse than the other semilocal functionals considered here.

\begin{center}
\noindent
\begin{table*}[htb]
\caption{Mean errors (ME), and mean absolute errors (MAE) of semilocal and hybrid functionals for the S22 test set and its subsets in kcal/mol}
\begin{tabular}{l cccccccccccc}
\hline\hline
	&\multicolumn{2}{c}{WI8}& &\multicolumn{2}{c}{MI7}& & \multicolumn{2}{c}{HB7}& & \multicolumn{2}{c}{S22} \\
\cline{2-3} \cline{5-6} \cline{8-9} \cline{11-12}
	&ME &MAE& &ME &MAE& &ME &MAE& &ME &MAE\\
\hline					
PBE	&4.8&4.8 & &2.0&2.0 & &1.1&1.1 & &2.8&2.8 \\
BLYP    &7.7&7.7 & &3.8&3.8 & &3.2&3.2 & &5.0&5.0 \\
BLYP*   &7.3&7.3 & &3.6& 3.6& & 2.8&2.8& &4.7&4.7 \\
TPSS	&6.0&6.0 & &2.7&2.7& & 1.9&1.9 & &3.7&3.7 \\
revTPSS	&4.3&4.3 & &2.4&2.4 & &2.0&2.0 & &2.9&2.9 \\
M06-L	&1.0&1.0 & &1.0&1.0& & 0.7&0.7 & &0.9&0.9\\
MGGA\_MS0	&3.0&3.0 & &1.2&1.2& & 0.8&0.8 & &1.7&1.7\\
MGGA\_MS1	&3.9&3.9& & 1.7&1.7 & &1.8&1.8 & &2.5&2.5\\
MGGA\_MS2	&3.0&3.0& & 1.3&1.3 & &1.4&1.4 & &1.9&1.9\\
\hline
PBEh	&4.6&4.6 & &1.8&1.8& &0.8&0.9 & &2.5&2.5\\
TPSSh	&5.8&5.8& & 2.6&2.6 & &1.8&1.8 & &3.5&3.5\\
revTPSSh	&5.0&5.0 & &2.3&2.3& & 1.8&1.8 & &3.1&3.1\\
B3LYP	&6.5&6.5 & &3.0&3.0& &2.0&2.0& & 4.0&4.0\\
M06	&1.4&1.4 & &0.9&0.9 & &0.7&0.7& & 1.0&1.0\\
MGGA\_MS2h	&3.0&3.0 & &1.3&1.3 & &1.2&1.2& & 1.9&1.9\\
\hline \hline
\end{tabular}

\label{table:S22}
\end{table*}
\end{center}

\begin{center}
\noindent
\begin{table*}[htb]
\caption{Mean errors (ME), and mean absolute errors (MAE) of semilocal and hybrid functionals with the D3 correction~\cite{GAEK_JCP_2010} for the S22 test set and its subsets in kcal/mol. All functionals use $s_8 = 0$ and $s_6 = 1$. See Ref.~\onlinecite{GAEK_JCP_2010} for the definitions of $s_6, s_8$, and $s_{r,6}$. }
\hspace*{-1.0cm}
\begin{tabular}{ccccccccccccccccccc}
\hline
\hline
 & \multicolumn{2}{c}{MGGA\_MS0+D3} & ~ & \multicolumn{2}{c}{MGGA\_MS1+D3} & ~ & \multicolumn{2}{c}{MGGA\_MS2+D3}              & & \multicolumn{2}{c}{MGGA\_MS2h+D3}            & &  \\
 & \multicolumn{2}{c}{$s_{r,6} = 1.15$} & ~ & \multicolumn{2}{c}{$s_{r,6} = 1.05$} & ~ &\multicolumn{2}{c}{$s_{r,6} = 1.14$} & ~ &\multicolumn{2}{c}{$s_{r,6} = 1.14$} & ~ & \\
            \cline{2-3}                       \cline{5-6}                                  \cline{8-9}                \cline{11-12}  
test set  & ~ ME ~ & ~MAE ~   & ~ & ~ ME ~ & ~ MAE ~ & ~ & ~ ME ~ & ~MAE ~                    & ~ & ~ ME  & ~ MAE              & ~ & ~ \\ 
\hline
G2/97     & -0.9  & 7.0 &  &  1.8  & 5.0 &  & -1.1  &  5.3    &  &  2.3  &   4.9  &  & \\ 
BH42/03   & -6.1  & 6.1 &  & -5.9  & 5.9 &  & -6.0  & 6.0    &  & -4.7  &   4.7  &  & \\ 
G3-3      & -7.4  & 13.1&  &  0.6  & 5.7 &  & -5.6  &  9.0    &  & -0.0  &  5.8  &  & \\ 
G3/99     & -3.1  & 9.1 &  &  1.4  & 5.3 &  & -2.6  &  6.6    &  &  1.6  &   5.2  &  & \\ 
HTBH38/04 & -5.8  &  5.9 &  & -5.8  &  5.8 &  & -6.0  & 6.0    &  & -4.6  &   4.6 &  & \\ 
NHTBH38/04& -6.5  & 7.0 &  & -6.5  &  6.9 &  & -7.1  &  7.4    &  & -5.4  &  5.8  &  & \\ 
\hline
WI8       & -0.0  & 0.4 &  & -0.2  &  0.4 &  & -0.2  &  0.3    &  & -0.1  & 0.3   &  & \\
MI7       & -0.3  &  0.3&  & -0.3  &  0.3 &  & -0.2  & 0.3    &  & -0.2  &  0.3   &  & \\
HB7       & -0.3  & 0.4 &  &  0.2  &  0.2 &  &  0.2  & 0.3    &  &  0.1  &  0.2   &  & \\
S22       & -0.2  & 0.3 &  & -0.1  & 0.3 &  & -0.0  & 0.3    &  & -0.1  &  0.2   &  & \\
\hline
\hline
\end{tabular} \\

\label{table:S22_D3}
\end{table*}
\end{center}

Ref. \onlinecite{KM_JCP_2012} presents a collection of different techniques for functionals sorted by the way
weak interactions are accounted for: step 1 (DFT-D~\cite{Grimme_JCC_2006}, etc), step 2 (DFT-D3~\cite{GAEK_JCP_2010}, vdW(TS)~\cite{TS_PRL_2009}, etc), step 3 (long-range density functionals~\cite{DRSLL_PRL_2004, VV_PRL_2009, KBM_JPCM_2010}), and step 4 and higher (RPA~\cite{Furche_PRB_2001}, etc).
Representative functionals of each group are tested for the S22 test set in Ref. \onlinecite{KM_JCP_2012}.
Note that the basis sets differ compared to our calculations, but the results can be compared roughly.
MGGA\_MS0 and MGGA\_MS2 yield results that are close to the vdW-DF~\cite{DRSLL_PRL_2004} which is on step 3 of this classification,
while MGGA\_MS1 fits best to the ground (below step 1). It should be noted that the classification is
methodological and not based on performance as some functionals in step 1 and 2 outperform some funtionals
on step 3 in Ref. \onlinecite{KM_JCP_2012}. 
Table~\ref{table:S22_D3} gives results of the MGGA\_MS functionals with the D3 correction of Grimme {\it et al}~\cite{GAEK_JCP_2010}. The parameters in the D3 correction term were optimized by fitting to the S22 data set for each functional. We found that the optimization is insensitive with respect to the parameter $s_8$, and thus set $s_8=0$. Then, there is only one fitting parameter $s_{r,6}$ left in the D3 correction term. $s_{r,6}$, which is the scaling factor of cutoff radii and determines the range of the dispersion correction inversely~\cite{GAEK_JCP_2010}, is 1.15, 1.05, 1.14, and 1.14 for MGGA\_MS0, MGGA\_MS1, MGGA\_MS2, and MGGA\_MS2h, respectively. All the four MGGA\_MS functionals with the D3 corrections deliver for the S22 data set similar MAEs around 0.3 kcal/mol, which are among the best of the D3-corrected functionals tested in Ref.~\onlinecite{GAEK_JCP_2010}. The smaller $s_{r,6}$ for MGGA\_MS1 indicates that MGGA\_MS1 captures less intermediate-range weak interactions than the other three do, as also shown in Table~\ref{table:S22}. Table~\ref{table:S22_D3} also shows that the D3 corrections have noticeable effects on the formation energies and the barrier heights. MGGA\_MS1+D3 is the best among the three MGGAs, while the hybrid MGGA\_MS2h+D3 improves further the formation energies and the barrier heights.

The functional pairs of global hybrid and its corresponding semilocal complement perform very similarly for
weak interactions on average.
The pair of BLYP* and B3LYP is the outlier, which might be related to the fact that B88 is too repulsive for rare gas dimers even when comparing to the exact exchange~\cite{KB_JCTC_2009}.
This general trend suggests that the performance of a hybrid of a semilocal functional inherits its performance for weak interactions
from its corresponding pure semilocal functional.
A possible explanation is that weak interactions are mainly correlation effects where exact exchange
does not contribute.
On the other hand, heats of formation, atomization energies, and barrier heights are significantly affected by correlation and 
exact exchange contributions.
However, in our constructions, MGGA\_MS2 is a byproduct of MGGA\_MS2h, which is fitted to the atomization energies of the G2/97 test set and the barrier heights of the BH42/03. Surprisingly, MGGA\_MS2h performs reasonably well for the S22 set, much better than the other four hybrids, namely, PBEh, TPSSh, revTPSSh, and B3LYP. Note the improvement of MGGA\_MS2h over the other four hybrids on the weak interactions is not a result of sacrifice of accuracy for atomization energies and barrier heights as shown in Table \ref{table:G3BH76}. The good performance of MGGA\_MS2h on the S22 set is then transfered to MGGA\_MS2.

When considering both Tables \ref{table:G3BH76} and \ref{table:S22} for the properties of atomization energies, barrier heights, and weak interactions, our three-parameter global hybrid MGGA\_MS2h overall is the second best hybrid after M06 that is heavily parameterized. Compared to MGGA\_MS2h, our semilocal functional MGGA\_MS1 performs surprisingly well for atomization energies of molecules and barrier heights of chemical reactions although it contains just one empirical parameter. However, the gain is obtained at the price of descriptions for weak interactions as demonstrated by the comparison with MGGA\_MS0 in Table \ref{table:S22}. This shows the limit of tuning only the $s-$dependence of the enhancement factor. Fortunately, by tuning the $\alpha-$dependence additionally, MGGA\_MS2 improves over MGGA\_MS0 for the atomization energies significantly while remaining comparable for the barrier heights and weak interactions.

\section{Conclusions}
\label{sect:conclusions}

By taking advantage of the understanding on the kinetic-energy-density dependence of MGGAs \cite{SXR_JCP_2012}, we construct a global hybrid meta-generalized gradient approximation (hybrid meta-GGA) with three empirical parameters (MGGA\_MS2h), which has robust performances on the molecular systems for the properties of heats of formation, barrier heights, and noncovalent interactions. The derived underlying MGGA\_MS2 improves over the original MGGA\_MS0 \cite{SXR_JCP_2012} for heats of formation significantly while retaining good performance for barrier heights and weak interactions. By relaxing the parameter $\kappa$ of MGGA\_MS0, we also obtained the one-parameter functional MGGA\_MS1 that performs surprisingly well for heats of formations and barrier heights.

\

\

\begin{acknowledgments}
Most calculations were done at Rice University.
This work was supported by the National Science Foundation under CHE-1110884 and
the Welch Foundation (C-0036).
RH thanks the Deutsche Forschungsgemeinschaft (DFG grant HA 5711/2-1).
JS, BX, and JPP acknowledge NSF support under Grant No. DMR-0854769, and Cooperative Agreement No. EPS-1003897, with additional support from the Louisiana Board of Regents. 
\end{acknowledgments}


\begin{thebibliography}{99}
\bibitem{fn1} COF$_2$ was included in the G2/97 and G3/99 test sets following Ref. \onlinecite{TPSSh}.
\end{thebibliography}
\end{document}